\documentclass[11pt]{article}
\usepackage[margin=1in]{geometry}
\usepackage{amsmath,amssymb,amsfonts}
\usepackage{amsthm}
\usepackage{algorithmic}
\usepackage{graphicx}
\usepackage{color}
\usepackage{textcomp}
\usepackage{float}
\usepackage{array}
\usepackage{pifont} 
\usepackage[justification=centering, font=footnotesize]{caption}
\usepackage[linesnumbered,ruled,vlined]{algorithm2e}
\usepackage{hyperref}
\usepackage{cite}

\usepackage{titlesec}
\titleformat{\section}{\bfseries\large}{\thesection.}{0.5em}{}
\titleformat{\subsection}{\bfseries\normalsize}{\thesubsection.}{0.5em}{}

\newtheorem{result}{\bf Result}

\newcommand{\brac}[1]{\left({#1}\right)}
\newcommand{\define}{\triangleq}
\newcommand{\lcm}{\text{LCM}}
\newcommand{\GR}{\text{GR}}
\newcommand{\GRM}{\text{GRM}}
\newcommand{\expect}[1]{{\bf E}\left[{#1}\right]}
\newcommand{\abs}[1]{\left\lvert{#1}\right\rvert}

\title{\bfseries Gauss-Ramanujan Functions: Constructions, Properties, and Applications in Communications and Signal Processing}

\author{\normalsize Sainath Bitragunta \\ 
	\small Electrical and Electronics Engineering Department \\
	\small BITS Pilani, Pilani Campus, Rajasthan-333031, INDIA \\
	\small \texttt{sainath.bitragunta@pilani.bits-pilani.ac.in}
}

\date{}

\begin{document}
	
	\maketitle

\begin{abstract}
	In this article, I construct a new set of functions based on Ramanujan sequences (RSEs),  Gaussian pulse (GP), and its delayed Gaussian pulse (DGP). The motivation for this construction is based on the special properties of RSEs, GP, and DGP. First, I present a procedure for constructing Gauss-Ramanujan (GauRam) functions using selected RSEs. I develop an insightful analysis for deterministic and stochastic overlap between GP and DGP. Specifically, I present exact and closed form approximation expressions for delay-averaged GP and DGP overlap and then evaluate them numerically. Later, I derive and analyze the mathematical (spectral) properties of selected GauRam functions. I extend the analysis by analyzing the Hilbert transform of the first-order GauRam function and validating orthogonality and its usefulness in analytic signal representations.
	Furthermore, I present insightful applications of these functions in communications and signal processing. Specifically, I present the continuous-wave Gauss-Ramanujan modulation (GRM) scheme, Gauss-Ramanujan Shift Keying (GRSK) scheme, and Gauss-Ramanujan wavelets and their analysis and comparisons with benchmarking. The desirable properties of these novel modulation schemes and wavelets enable their use in next-generation hybrid and energy-efficient communication systems and signal processing.
\end{abstract}

\section*{\normalsize Keywords}
\small 
Applications, Benchmarking, Continuous-wave Modulation, Delayed-Gaussian Pulse, Digital modulation, Gaussian Pulse, Gauss-Ramanujan functions, Gauss-Ramanujan Wavelet, Spectral Properties, Ramanujan sequences, Performance analysis.


\maketitle

\section{INTRODUCTION AND MOTIVATION}

Designing novel mathematical functions with desirable properties and investigating their properties in different domains is essential in advancing analog/digital modulation techniques and signal processing tools, such as developing new mother wavelets. Classical pulses such as Gaussian pulse, sinc, and polynomials (for example, Butterworth and Chebyshev) have been extensively utilized in various applications, ranging from communications and signal processing~\cite{haykin2014DigComSys,oppenheim1999discrete}. 

With the increasing demand for more efficient and flexible signal sets, researchers explored new constructions to develop novel function families that exhibit desirable properties. These include orthogonality, smoothness, dispersion-free nature, distortion-free nature, and well-controlled spectral properties. The construction of such functions enhances existing analog and digital signal processing tools and enables new approaches in modulation schemes, efficient adaptive filtering, and multi-resolution analysis~\cite{Atlas1999SPmag}.

The main idea behind novel continuous functions construction lies in integrating Ramanujan sequences with a well-established Gaussian pulse and its delayed variants. Traditional modulation techniques rely on discrete deterministic/random information sequences and pulse shaping filters. By systematically embedding discrete Ramanujan sequences into continuous Gaussian pulse variants, it is possible to leverage the benefits of both domains— i) combining the localization and structured properties of discrete sequences with the smoothness and ii) frequency domain advantages of continuous functions. 

This hybrid approach allows for developing novel Gauss-Ramanujan functions with desirable properties, including higher flexibility and greater adaptability for specific application requirements (for example, classical and quantum communications and signal processing). Studying the mathematical properties of Gauss-Ramanujan functions significantly contributes to the rich fields of signals, systems, signal processing, and communications.

Below, I present the relevant literature that motivated this work. I consider two perspectives, namely, communications and signal processing.

{\em Communications Perspective:} In the era of fifth-generation (5G) and beyond—including 5G-enabled IoT scenarios—there is a growing need for robust modulation schemes that can handle high mobility, Doppler effects, and delay spreads. Orthogonal time frequency space (OTFS) modulation has emerged as a powerful candidate by mapping symbols in the delay-Doppler domain and offering superior performance under time-varying conditions~\cite{hadani2017orthogonal, raviteja2018interference}. However, the choice of pulse shape critically influences system performance, particularly in terms of interference management and spectral efficiency.

The proposed modulation scheme, Gauss-Ramanujan shift keying (GRSK), introduces a new dimension in pulse shaping, leveraging delay diversity through the tunable delay parameter $T_0$. This pulse shape structure allows adaptive symbol placement and improved resilience in doubly dispersive channels. It effectively exploits delay diversity and enhances delay-Doppler sparsity, both advantageous for OTFS-based communication systems in dynamic environments~\cite{surabhi2019optimal, Aldababsa2024otfs}.

{\em Signal Processing Perspective:} From a waveform design standpoint, GRSK can be interpreted as a generalized wavelet-like structure with Gaussian atoms weighted by Ramanujan sequences. Compared to Hermite wavelets—which are fixed and lack tunable parameters—GRSK provides superior adaptability via its configurable delay shifts (integer multiples of delay parameter $T_0$). This adaptability enhances the time-frequency localization and supports better interference suppression. Additionally, GRSK’s structure aligns with modern requirements for compressive sensing and sparse signal recovery, making it highly relevant for communication systems and emerging signal processing frameworks~\cite{mallat1999wavelet, duarte2011structured}.

Both the above perspectives justify motivation for the exploration of novel modulation schemes for next generation mobile wireless communication systems and Gauss-Ramanujan wavelets for novel and efficient signal processing. 

\subsection{Novelty and Key Contributions}

Based on the literature review, the design of such novel continuous-time functions (Gauss-Ramanujan functions) and their applications have not been explored in the literature. The construction is non-trivial, the properties are interesting and insightful, and the applications in next generation heterogeneous communication systems and signal processing seem promising. 

Some of the research gaps identified in the literature include the lack of waveforms offering delay diversity and enhancing delay-Doppler sparsity and the limited exploration of delay-adaptable wavelets with desirable time-frequency containment. Addressing these gaps, I propose novel GRSK modulation and Gauss-Ramanujan wavelets, which are advantageous for next generation heterogeneous communication and signal processing systems. I summarize my contributions below:

\begin{itemize}
	\item \textbf{GauRam functions construction and near-orthogonality investigation:} I propose a simple and novel construction of \textit{Gauss-Ramanujan (GauRam)} functions by combining Gaussian pulse and its delayed variant with weights based on Ramanujan sequences. This construction introduces novel pulse shapes that are analytically tractable and suitable for waveform design in spectrum resource and interference-constrained 5G and beyond communication systems.
	
	\item \textbf{Time and frequency domain properties and analysis:} I obtain more insights into the behavior of GauRam functions by deriving several properties of first-order GauRam functions in both the time and frequency domains. These representations reveal the flexible spectral shaping capabilities of the GauRam functions, making them candidates for interference-resilient modulated waveform designs.
	
	\item {\textbf{Analysis for first-order GauRam functions:}} A thorough mathematical analysis is provided for the first-order GauRam function. The key analytical results include
	\begin{itemize}
		\item {\em Near-orthogonality condition:} I derive a closed form expression for the delay to ensure arbitrarily small overlap and near-orthogonality. This condition also helps to ensure minimal inter-symbol interference.
		\item \textit{Hilbert transform analysis:} I derive a closed form expression for the Hilbert transform of the first-order GauRam function and present an insightful plot to visually justify the orthogonality between the GauRam function and its Hilbert transform.
		\item \textit{Offset-based stochastic delay modeling:} I model the delay parameter in the first-order GauRam function as a uniform random variable. For it, I derive both exact and approximate closed-form expressions for the mean overlap as a function of delay offset.
	\end{itemize}
	
	\item \textbf{Applications in modulation waveform and pulse shape design:} I apply the proposed GauRam functions in the design of a continuous-wave \textit{Gauss-Ramanujan Modulation (GRM)} scheme and introduce a new Gauss-Ramanujan Shift Keying (GRSK) scheme. The normalized power spectral density (PSD) of the GRSK pulse is compared with the benchmarking scheme Gaussian Minimum Shift Keying (GMSK) pulse. The comparative analysis is useful in leveraging potential benefits such as delay diversity and adaptive pulse shaping for efficient interference management and mitigation.
	
	\item \textbf{Comparative Wavelet Analysis:} I compare the proposed GauRam-based wavelet (first-order) to Hermite wavelet (first-order), demonstrating improved adaptability due to the tunable delay parameter, with potential applications in adaptive wavelet-based efficient signal processing and feature extraction.
	
\end{itemize}


\section{MATHEMATICAL PRELIMINARIES: RAMANUJAN MEETS GAUSS}
I first present a brief discussion on Ramanujan sums and the finite sequences comprising the sums. Later, I discuss continuous-time Gaussian pulse (GP) and delayed Gaussian pulse (DGP).

\subsection{RAMANUJAN SUMS AND SEQUENCES}

Let $k$ and $R$ denote positive integers such that $k$ and $R$ are coprime (or relatively prime). That is, $\gcd(k,R) = 1$. Note that coprimes do not have a common factor greater than $1$. Let $\mathbb{S}_{R}(m)$ denote the Ramanujan sum (RSU). Mathematically, the $R^\text{th}$ RSU ($R \geq 1$) is defined as follows~\cite{PPV2015Eusipco}.
\begin{equation}
	\label{eq:RSU_def}
	\mathbb{S}_{R}(m) = \sum_{{k = 1},\,(k, R) = 1}^{R} e^{j 2\pi \frac{k}{R} m}.
\end{equation}
It can be shown that $\mathbb{S}_{R}(m)$ is periodic with period $R$. That is, $\mathbb{S}_{R}(m + R) = \mathbb{S}_{R}(m)$. 

{\em Orthogonality:} Let $R_{1} \neq R_{2}$. For RSEs, I can verify the following property~\cite{PPV2015Eusipco}:
\begin{equation}
	\label{eq:Ortho_pty_RSUs}
	\sum_{{n = 0}}^{\lcm\{R_{1}, R_{2}\}-1} \mathbb{S}_{R_{1}}(n) \mathbb{S}_{R_{2}}(n) = 0.
\end{equation}

{\em Ramanujan Sequence (RSEs):} To construct RSEs, I consider normalizing each component with $\mathcal{L}^{2}$-norm. $\tilde{\mathbb{S}}_{R}(m), R > 1$ can be termed as a sequence. Some RSEs are: $\{\frac{1}{\sqrt{2}}, -\frac{1}{\sqrt{2}}\}, \{1,-\frac{1}{2},-\frac{1}{2}\}, \{\frac{1}{\sqrt{2}},0,-\frac{1}{\sqrt{2}},0\}$. Note that the first and the third RSEs have unit $\mathcal{L}^{2}$-norm. In this article, I consider the three RSEs to construct Gauss-Ramanujan (GauRam) functions and study their properties. However, I can extend the sequence length further as required.

For the construction of GauRam functions, I use the RSE components as weights for GP and DGP. Below, I discuss GP and DGP and state their properties.

\subsection{GP and Delayed-GP: Near Orthogonality}

Consider the normalized GP $g(t) = e^{-\pi t^2}, t \in \mathbb{R}$. It can be verified that the area under the normalized GP is unity, that is~\cite{haykin2014DigComSys},
\begin{equation}
	\label{eq:GP_area}
	\int_{-\infty}^{\infty} e^{-\pi t^2} \,dt = 1.
\end{equation}

Another interesting property of the GP is that its continuous-time (CT) Fourier transform (CTFT) is itself. That is, $e^{-\pi t^2} \Leftrightarrow e^{-\pi f^2}$. Let $g(t - T_0) = e^{-\pi \brac{t-T_0}^2}$, where $T_0 > 0$ denotes finite delay. From the time-shifting property, I have $e^{-\pi \brac{t-T_0}^2} \Leftrightarrow e^{-\pi f^2} e^{-j 2\pi f T_0}$. Applying Rayleigh's energy theorem~\cite{haykin2014DigComSys}, it can be shown that the area under the DGP in $f$-domain is also unity.

I consider GP and DGP and quantify the overlap between the two. Specifically, I determine the condition for arbitrarily small overlap $\epsilon$. Let $h(t) = e^{-\pi \brac{t-T_0}^2}, t \in \mathbb{R}$. 

The inner product of GP and DGP is defined as
\begin{equation}
	\langle g(t), h(t) \rangle = \int_{-\infty}^{\infty} g(t) \, h(t) \, dt 
	= \int_{-\infty}^{\infty} e^{-\pi t^2} \, e^{-\pi (t-T_0)^2} \, dt.
\end{equation}

After simplifying the integral, I obtain a closed-form expression for the inner product, as shown below.
\begin{equation}
	\label{eq:gh_ip}
	\langle g(t), h(t) \rangle = \frac{1}{\sqrt{2}}\exp\left(-\frac{\pi T_0^2}{2}\right).
\end{equation}
Since $\sigma = \frac{1}{\sqrt{2\pi}}$, I have
\begin{equation}
	\langle g(t), h(t) \rangle = \frac{1}{\sqrt{2}}\exp\left(-\frac{ T_0^2}{4 \sigma^2}\right).
\end{equation}

Below, I state an insightful result on the condition on delay parameter $T_0$ for arbitrarily small ($\epsilon$) overlap. 
\begin{result}
	\label{result:eps_overlap}
	Define
	\begin{equation}
		\langle g(t), h(t) \rangle = \epsilon,
	\end{equation}
	where \(0 < \epsilon \ll 1\) (for example, \(\epsilon = 1\% \)). The delay 
	\begin{equation}
		\label{eq:GP_DGP_time_spacing}
		T_0 (\sigma, \epsilon) = \sqrt{-4 \sigma^2 \ln\left(\sqrt{2}\,\epsilon\right)} = 2\sigma \sqrt{-\ln\left(\sqrt{2}\,\epsilon\right)},
	\end{equation}
	where $\sigma = \frac{1}{\sqrt{2\pi}}$.
\end{result}
\begin{proof}
	From~\eqref{eq:gh_ip}, I have
	\begin{equation}
		\nonumber
		\frac{1}{\sqrt{2}}\exp\left(-\frac{\pi T_0^2}{2}\right)=\epsilon.
	\end{equation}
	Taking the natural logarithm on both sides, I get
	\begin{equation}
		-\frac{\pi T_0^2}{2} = \ln\left(\sqrt{2}\,\epsilon\right)
		.
	\end{equation}
	Simplifying further, I get
    \begin{equation}
		 T_{0}^{2} = -\frac{2}{\pi}\ln\left(\sqrt{2}\,\epsilon\right).
	\end{equation}
	Therefore, the delay \(T_0\) required to achieve an arbitrarily small overlap of \(\epsilon\) is given by
	\begin{equation}
		T_0 = \sqrt{-\frac{2}{\pi}\ln\left(\sqrt{2}\,\epsilon\right)}.
	\end{equation}
	By substituting $\sigma = \frac{1}{\sqrt{2\pi}}$, I get the desired result.
	%
\end{proof}



{\em Remarks:} Below, I state the following useful remarks.
\begin{itemize}
	\item The inner product \(\langle g(t), h(t) \rangle\) measures the overlap between the GP and its delayed version. As \(T_0\) increases, the overlap decays exponentially.
	
	\item For a very small overlap (e.g., \(\epsilon = 0.01\)), the above formula quantifies how large \(T_0\) must be relative to the dispersion parameter ($\sigma$) of the GP.
	
	\item {\em $6\sigma$-rule:} The $6\sigma$-rule is a rule of thumb that indicates that almost all the energy of a Gaussian is contained within $\pm 3\sigma$. Therefore, if two GPs are separated by $6\sigma$, their overlapping tails contribute very little energy. For example, let $T_0 = 6 \sigma \approx 2.4$. I compute $\epsilon = \frac{1}{\sqrt{2}}\exp\left(-\frac{\pi T_0^2}{2}\right) = 8.3 \times 10^{-5}$. The lower value of $\epsilon$ indicates that GP and DGP become approximately orthogonal when $T_0$ exceeds $6\sigma \approx 2.4$. Fig.~\ref{fig:GP_DGP_sep_6sig} illustrates this.

	\begin{figure}[h!]
		\centerline{    \includegraphics[width=1\linewidth]{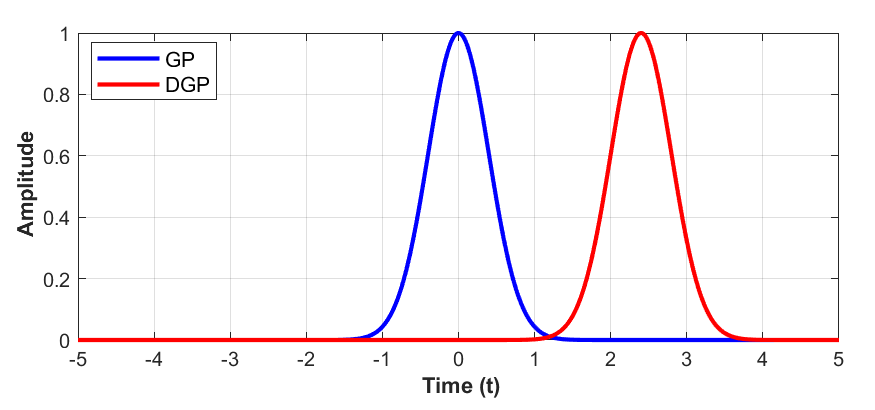}} 
		\caption{An illustration of $6\sigma$ separation between GP and DGP.}
		\label{fig:GP_DGP_sep_6sig}
	\end{figure}

	\item The result on $T_0$~\eqref{eq:GP_DGP_time_spacing} is significant in applications such as multi-carrier communications where near-orthogonality between time-shifted pulses is desired.
	
	\subsubsection{Stochastic Modeling of Delay in DGP}
	
	Imperfections in physical realization could cause an offset $\delta$ in $T_{0}$. Let $\tau_{0}$ represent a uniformly distributed random variable. Suppose $\tau_{0} \sim \mathcal{U}[T_{0} - \delta, \, T_0 + \delta], \, 0 \leq \delta << T_{0}$. 
	
	Below, I state an interesting analytical result on the average amount of overlap as a function of offset $\delta$. The result comprises exact and approximate expressions for the delay-averaged GP and DGP overlap.

 \begin{result}
 	\label{result:mean_overlap}
 	
Let $\overline{\epsilon}$ denote the average overlap of GP and DGP. The exact delay-averaged GP and DGP overlap can be expressed as
 \begin{equation}
 	\label{eq:av_olap}
 	\overline{\epsilon} = \frac{Q\brac{\sqrt{\pi}(T_{0} - \delta)} - Q\brac{\sqrt{\pi}(T_{0} + \delta)}}{2 \delta},
 \end{equation}
 where $Q(.)$ denotes Gaussian $Q$-function defined on the standard Gaussian distribution~\cite{haykin2014DigComSys}.
 
 The closed form approximation for the delay-averaged GP and DGP overlap is given by
 \begin{equation}
 	\label{eq:Av_ol_clf}
 	\overline{\epsilon} \approx \frac{1}{\delta}\exp\Bigl[-\Bigl(\alpha\pi T_0^2 + \alpha \pi\delta^2 + \beta \sqrt{\pi}T_0 + \gamma \Bigr)\Bigr] \sinh\Bigl(2\alpha \pi T_0\delta + \beta \sqrt{\pi}\delta\Bigr),
 \end{equation}
 where $\alpha = 0.3842$, $\beta = 0.7640$, and $\gamma = 0.6964$, which are chosen to minimize sum of squared errors.
 \end{result}
 \begin{proof}
 	Consider the model: $\tau_{0} \sim \mathcal{U}[T_{0} - \delta, \, T_0 + \delta]$, which is a uniform random variable, the instantaneous overlap $\epsilon$, which a function of stochastic delay $\tau_{0}$, is also a random variable.
 	
 	From~\eqref{eq:gh_ip}, the delay-averaged GP and DGP overlap can be expressed as
 	\begin{equation}
 		\label{eq:av_olap_def}
 		\overline{\epsilon} = \expect{\frac{1}{\sqrt{2}}\exp\left(-\frac{\pi \tau_{0}^2}{2}\right)}.
 	\end{equation}
 	%
 	
 	
 The probability density function (pdf) of $\tau_0$ is given by
 	\begin{equation}
 		p_{\tau_0}(\tau) = \frac{1}{2\delta}, \quad \tau \in [T_{\ell}, T_{u}],
 	\end{equation}
 where $T_{\ell} = T_0 - \delta$, $T_{u} = T_0 + \delta$.
 Therefore, 
 	\begin{equation}
 		 = \frac{1}{2\delta\sqrt{2}} \int_{T_{\ell}}^{T_{u}} \exp\left(-\frac{\pi \tau^2}{2}\right) d\tau.
 	\end{equation}

Using substitution: \( v = \sqrt{\frac{\pi}{2}} \tau \Rightarrow dv = \sqrt{\frac{\pi}{2}}\, d\tau \). The lower and upper limits become $\sqrt{\frac{\pi}{2}} T_{\ell}$, $ \sqrt{\frac{\pi}{2}} T_{u}$. The single integral can be expressed as
 	\begin{equation}
 		\frac{1}{2\delta\sqrt{2}}  \frac{1}{\sqrt{\frac{\pi}{2}}} \!\int_{\sqrt{\frac{\pi}{2}} T_{\ell}}^{\sqrt{\frac{\pi}{2}} T_{u}} e^{-v^2}\,dv
 		\!=\! \frac{1}{2\delta\sqrt{\pi}} \!\int_{\sqrt{\frac{\pi}{2}} T_{\ell}}^{\sqrt{\frac{\pi}{2}} T_{u}} e^{-v^2}\, dv.
 	\end{equation}
 	%
 	
Consider the following standard integral:
 	\begin{equation}
 		\int_{L}^{U} e^{-t^2} dt = \sqrt{\pi} \left[ Q\left(\sqrt{2}L \right) - Q\left(\sqrt{2}U\right) \right].
 	\end{equation}
Using the above integral and simplifying, I get
 	\begin{equation}
 		\overline{\epsilon} = \frac{1}{2\delta} \left[ Q\left(\sqrt{\pi} T_{\ell}\right) - Q\left(\sqrt{\pi} T_{u}\right) \right].
 	\end{equation}

 Substituting $T_{\ell} = T_0 - \delta$, $T_{u} = T_0 + \delta$, I get the desired result for delay-averaged GP and DGP overlap.
%
%

{\em Close form approximation:}

 	Consider the following approximation~\cite{Lopez2011ieeeTcom}:
 	\begin{equation}
 		Q(y) \approx \exp\left( - (\alpha y^2 + \beta y + \gamma) \right),
 	\end{equation}
 	Let
 	\begin{equation}
 		y_{\pm} = \sqrt{\pi}(T_0 \pm \delta),
 	\end{equation}
 	Therefore, 
 	\begin{equation}
 		Q(y_{\pm}) \approx \exp\left( -\left[ \alpha y_{\pm}^2 + \beta y_{\pm} + \gamma \right] \right).
 	\end{equation}

 	The delay-averaged GP and DGP overlap is approximately given by
 	\begin{equation}
 		\overline{\epsilon} \approx \frac{1}{2\delta} \\ \times \left[ \exp\left(-\alpha y_-^2 - \beta y_- - \gamma \right) 
 		- \exp\left(-\alpha y_+^2 - \beta y_+ - \gamma \right) \right] 
 		= \frac{\mathcal{M}}{2\delta} \left[ \exp(Y) - \exp(-Y) \right] = \frac{\mathcal{M}}{\delta} \sinh(Y),
 	\end{equation}
 	where
 	\begin{align}
 		\mathcal{M} &= \exp\left( -\brac{\alpha \pi T_0^2 + \beta \pi \delta^2 + \beta \sqrt{\pi} T_0 + \gamma} \right), \\
 		Y &= 2 \alpha \pi T_0 \delta + \beta \sqrt{\pi} \delta.
 	\end{align}
 	
  	
 \end{proof}
 
	

{\em Remarks on the accuracy of approximation:}
\begin{figure}[h!]
	\centerline{    \includegraphics[width=1\linewidth]{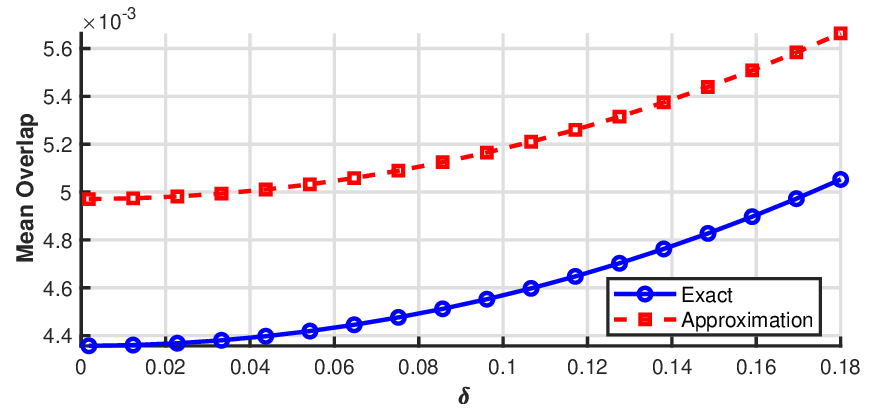}} 
	\caption{Mean overlap as a function of $\delta$: Exact versus closed form approximation.}
	\label{fig:Avg_overlap_exact_vs_approx_v1}
\end{figure}

Figure~\ref{fig:Avg_overlap_exact_vs_approx_v1} shows the mean overlap as a function of $\delta$. I observe that for $\delta = 0.09$ (that is, $5$\% of $T_{0}$), the closed-form approximation has a percent error of roughly $2.6$\% compared to the exact expression.

\end{itemize}

As mentioned earlier, the novelty in my work lies in the construction of GauRam functions, their spectral properties, and applications in communications and signal processing. The delay parameter $T_0$ plays a key role in all the GauRam functions. Below, I present the construction of GauRam functions using RSEs, GP, and DGP. 

\section{GAURAM FUNCTIONS: CONSTRUCTIONS AND ILLUSTRATIONS}

We define the \textit{Gauss-Ramanujan function} of order $q$ as:
\begin{equation}
	\label{eq:grf_gen}
	\GR_{R}(t) = \sum_{k=0}^{R-1} w_{k}^{(R)} g(t - k T_0),
\end{equation}
where $w_{k}^{(R)}$ are the normalized weights from the Ramanujan sum $\mathbb{S}_R(n)$.

The zeroth order GauRam function is the basic GP, that is, $\GR_{0}(t) = g(t) = e^{-\pi t^2}$. Consider two element RSE, that is,  $\{\frac{1}{\sqrt{2}}, -\frac{1}{\sqrt{2}}\}$. Consider $g(t) = e^{-\pi t^2}, t \in \mathbb{R}$ and $h(t) = e^{-\pi \brac{t-T_0}^2}, t \in \mathbb{R}$. The first order GauRam functions is given by $\frac{1}{\sqrt{2}} g(t) - \frac{1}{\sqrt{2}} h(t)$. In terms of GP $g(t)$, 
$\frac{1}{\sqrt{2}} g(t) - \frac{1}{\sqrt{2}} g(t - T_0) \define \GR_{I}(t; T_0)$.

\begin{figure}[h!]
	\centerline{    \includegraphics[width=1\linewidth]{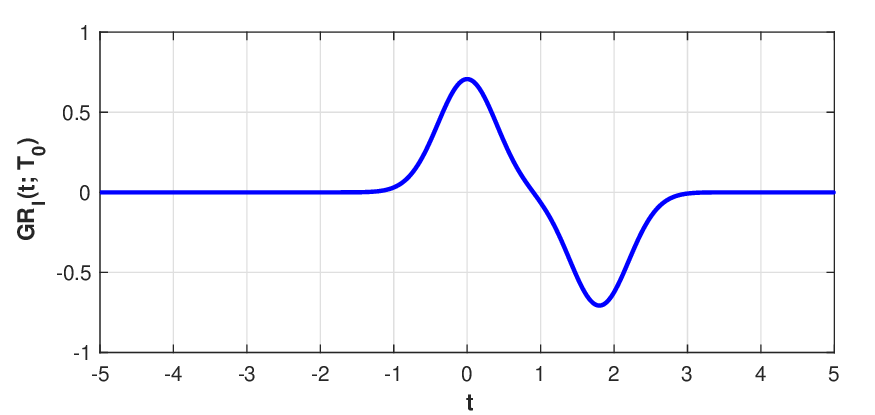}} 
	\caption{Illustration of $\GR_{I}(t; T_0)$ ($T_0 = 1.8$).}
	\label{fig:GP_formOne_Illu_v1}
\end{figure}

Therefore, the first-order GauRam function is given by 
\begin{align}
	\label{eq:GauRam_FO_func}
	\GR_{I}(t; T_0) &= \frac{1}{\sqrt{2}} g(t) - \frac{1}{\sqrt{2}} g(t - T_0), \\ 
	&= \frac{1}{\sqrt{2}} e^{-\pi t^2} - \frac{1}{\sqrt{2}}  e^{-\pi \brac{t-T_0}^2}, t\in \mathbb{R}.
\end{align}
The first-order GauRam function $\GR_{I}(t; T_0)$ is plotted in figure~\ref{fig:GP_formOne_Illu_v1}. 

\begin{figure}[h!]
	\centerline{\includegraphics[width=1\linewidth]{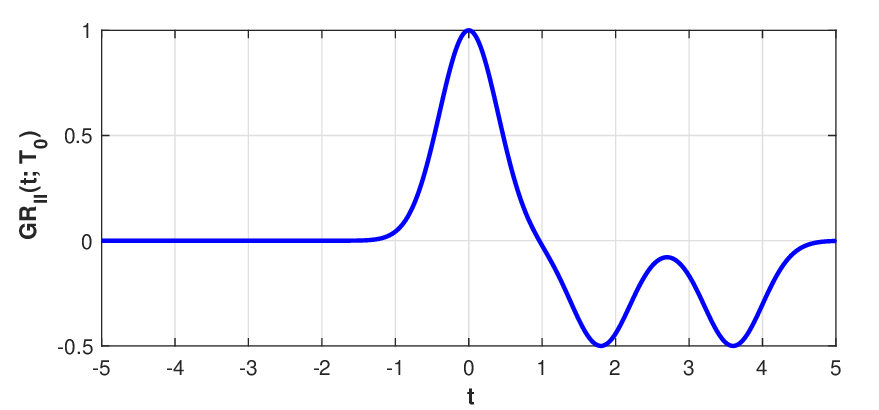}} 
	\caption{An illustration of $\GR_{II}(t; T_0)$ ($T_0 = 1.8$).}
	\label{fig:GP_formTwo_Illu_v1}
\end{figure}

The second-order GauRam function is given by 
\begin{align}
	\label{eq:GauRam_Sec_Order}
	\GR_{II}(t; T_0) &=  g(t) - \frac{1}{2} g(t - T_0) - \frac{1}{2} g(t - 2T_0), \\
	&= e^{-\pi t^2} - \frac{1}{2}  e^{-\pi \brac{t-T_0}^2} - \frac{1}{2}  e^{-\pi \brac{t-2T_0}^2}, t\in \mathbb{R}.
\end{align}
The first-order GauRam function $\GR_{II}(t; T_0)$ is plotted in~\ref{fig:GP_formTwo_Illu_v1}. 

{\em Remarks:} 
\begin{itemize}
	
	\item Using the above approach, one can construct higher-order (third-order, fourth-order, and above) GauRam functions. For instance, the third-order GauRam function is given by
	\begin{align}
		\label{eq:GauRam_Third_Order}
		\GR_{III}(t; T_0) &= \frac{1}{\sqrt{2}} g(t) - \frac{1}{\sqrt{2}} g(t - 2 T_0) , \\ 
		&= \frac{1}{\sqrt{2}} e^{-\pi t^2} - \frac{1}{\sqrt{2}}  e^{-\pi \brac{t-2T_0}^2}, t\in \mathbb{R}.
	\end{align}
	%
	
	
\end{itemize}

In the subsequent section, I investigate the GauRam functions' properties in the $f$-domain. 
For ease of analysis, I focus on the first‐order "GauRam" functions ($\GR_{I}(t; T_0)$), whose construction is relatively simple. However, their properties and the analysis are nontrivial and insightful.

\section{MATHEMATICAL PROPERTIES AND REMARKS}

I first derive the magnitude and phase responses of the first order GauRam function $\GR_{I}(t; T_0)$ and present useful remarks. Later, I consider the Hilbert transform (HT) of the $\GR_{I}(t; T_0)$. 

\subsection{First-Order GauRam Function: Frequency Response}

Note that GauRam functions are real-valued, continuous, and differentiable functions. To develop insightful mathematical properties, I first consider $\GR_{I}(t; T_0)$ and its Fourier transform $\GR_{I}(f; T_0)$. Using the properties of CTFT, it is straightforward to show that 
\begin{align}
	\label{eq:GauRam_I_fdom}
	\GR_{I}(f; T_0) &= \frac{1}{\sqrt{2}} e^{-\pi f^2} - \frac{1}{\sqrt{2}}  e^{-\pi f^2} e^{-j2\pi f T_0},\\
	&= \frac{G(f)}{\sqrt{2}} (1-e^{-j2\pi f T_0}),
\end{align}
where $G(f) = e^{-\pi f^2}$. 

\begin{figure}[h!]
	\centerline{    \includegraphics[width=1\linewidth]{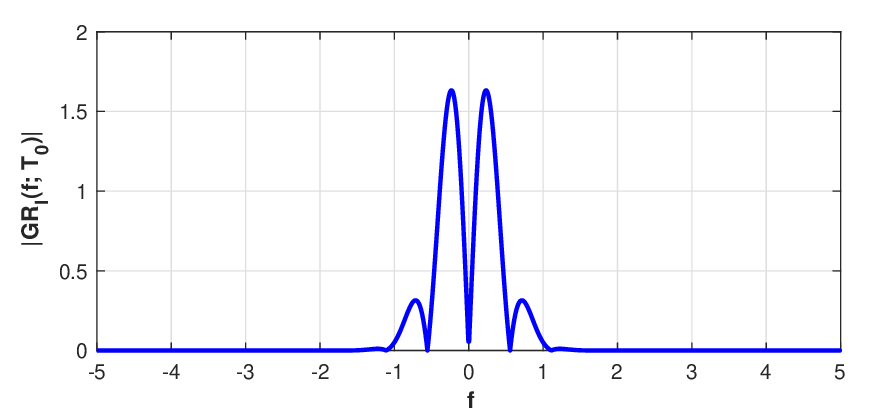}} 
	\caption{Magnitude spectrum of $\GR_{I}(t; T_0)$ ($T_0 = 1.8$).}
	\label{fig:GP_formOne_MS_Illu_v1}
\end{figure}

Furthermore, I find that the magnitude spectrum (MS) is given by
\begin{align}
	\label{eq:GauRam_I_MagSpec}
	\abs{\GR_{I}(f; T_0)} &= G(f) \sqrt{1 - \cos(2\pi f T_0)}, \\ 
	&= 2 G(f) \abs{\sin\brac{\pi f T_0}}.
\end{align}
A representative plot of the MS is shown in~\ref{fig:GP_formOne_MS_Illu_v1}.

{\em Remarks:} i) I see that $	\abs{\GR_{I}(f; T_0)}$ has zero crossing at $f = \frac{m}{T_0}, m \in \mathbb{Z}$. 

ii) The phase response $\psi(f)$ is given by 
\begin{equation}
	\label{eq:GauRam_I_PhaseRes}
	\psi(f) = -\pi f T_0 + \frac{\pi}{2}.
\end{equation}
Therefore, I have an affine phase response due to the additional phase of $\frac{\pi}{2}$.

iii) Since the phase response is affine, the issue of phase distortion is absent. However, it does impose a phase offset of $\frac{\pi}{2}$ across all frequencies. 

iv) {\em Group delay $\tau_{g}(f)$:} It is easy to show that the group delay is $\pi T_0$. Suppose $T_0$ is fixed. The group delay is constant. Therefore, dispersion and phase distortion are absent. 

These properties are desirable and useful for designing novel and efficient modulation techniques and new wavelet constructions. The spectral analysis can be extended  for other GauRam functions $\GR_{II}(t; T_0)$, $\GR_{III}(t; T_0)$ and so on. I present some interesting applications of GauRam functions in the subsequent section.

\subsection{First-Order GauRam Function: Hilbert Transform}

The HT is fundamental for constructing analytic signals and quadrature components in communications and signal processing~\cite{bracewell}. Composite pulses formed by time‐shifted Gaussians can approximate orthonormal basis functions in ultra‐wideband and radar applications~\cite{boashash}. Furthermore, HT is useful in representing single-sideband (SSB) continuous-wave modulated signals. 

First, I state the result on the HT of $g(t) = \exp(-\pi t^2)$ and prove it. Later, using this result and properties of the Fourier transform, I present an analytical result on the HT of $\GR_{I}(t ; T_0)$ and its proof.

\begin{result}
	\label{result:gt_ht}
	Let $\hat{g}(t)$ denote the HT of $g(t)$. 
	The Hilbert transform of the GP (the zeroth order GauRam function) is given by~\cite{King2009}
	\begin{equation}
		\hat{g}(t) = \frac{2}{\sqrt{\pi}} \mathbb{D}_{+}(\sqrt{\pi} t),
	\end{equation}
	where $\mathbb{D}_{+}(.)$ denotes the Dawson function given by
	\begin{equation}
		\mathbb{D}_{+}(z) = e^{-z^2} \int_{0}^{z} e^{u^2} du.
	\end{equation}

\end{result}
\begin{proof}
	
	In the time-domain, $\hat{g}(t) = \frac{1}{\pi t} \circledast g(t)$, where $\circledast$ denotes the convolution. The function $\hat{g}(t)$, which is the HT of $g(t)$ has Fourier transform
	\begin{equation}
		\hat{G}(f) = -j \, \mathrm{sgn}(f) \, G(f) =  -j \, \mathrm{sgn}(f) e^{-\pi f^2},
	\end{equation}
	where $\mathrm{sgn}(f)$ denotes the signum function.
	
	To find the HT of $g(t)$, I need to find the inverse Fourier transform of $\hat{G}(f)$. Let $\mathcal{F}^{-1}$ denote the inverse Fourier transform.  
	The HT of $g(t)$ is given by
	\begin{align}
		\hat g(t)
		&= \mathcal{F}^{-1}\{-j\,\mathrm{sgn}(f)e^{-\pi f^2}\}(t)\nonumber\\
		&= -j\int_{-\infty}^\infty \mathrm{sgn}(f)\,e^{-\pi f^2}e^{j2\pi f t}\,df.
	\end{align}

	Applying the definition of $\mathrm{sgn}(f)$ for $f>0$ and $f<0$ yields
	\begin{equation}
		\hat g(t)
		= -j\Bigl[\int_{0}^{\infty}e^{-\pi f^2}e^{j2\pi ft}df
		-\!\int_{0}^{\infty}e^{-\pi f^2}e^{-j2\pi ft}df\Bigr]
	\end{equation}
	Let $\overline{\mathcal{I}}$ denote the complex conjugate of $\mathcal{I}$.  Simplifying further, I get
	\begin{align}
		\hat g(t)
		&= -j\Bigl[\underbrace{\int_{0}^{\infty}e^{-\pi f^2}e^{j2\pi ft}\,df}_{\mathcal{I}(t)}
		\;-\;\underbrace{\int_{0}^{\infty}e^{-\pi f^2}e^{-j2\pi ft}\,df}_{\overline{\mathcal{I}(t)}}\Bigr]
		\label{eq:hg_I_diff}\\
		&= -j\bigl[\mathcal{I}(t) - \overline{\mathcal{I}(t)}\bigr]
		= -j\bigl[2j\,\Im\{\mathcal{I}(t)\}\bigr]
		= 2\,\Im\{\mathcal{I}(t)\},
		\label{eq:hg_Im}
	\end{align}
	where $\Im(.)$ denotes imaginary part.

	Define
	\[
	\mathcal{I}(t)=\int_{0}^{\infty}e^{-\pi f^2}e^{j 2\pi ft}df.
	\]
	Using the following standard integral for $\Re(p)>0$~\cite{abramowitz_stegun},
	\[
	\int_{0}^{\infty}e^{-p f^2 + q f}\,df
	=\tfrac12\sqrt{\tfrac{\pi}{p}}\,
	\exp\!\bigl(\tfrac{q^2}{4p}\bigr)
	\text{erfc}\!\bigl(-\tfrac{q}{2\sqrt{p}}\bigr),
	\]
	with $p = \pi$, $q = j2\pi t$, I get
	\[
	\mathcal{I}(t)
	=\tfrac12\,e^{-\pi t^2}\,
	\text{erfc}\!\bigl(-j\sqrt{\pi}\,t\bigr).
	\]
	
	I use the following facts:  $\text{erfc}(-j\nu)=1+ j\,\text{erfi}(\nu)$ and $\text{erfi}(\nu)=\tfrac{2}{\sqrt\pi}\!\int_0^{\nu} e^{s^2}\,ds$, where $\text{erfc}(.)$ is the complementary error function and $\text{erfi}(.)$ is the imaginary error function.  
	
	Using the above fact, I find that 
	\[
	\Im\{\mathcal{I}(t)\}
	=\tfrac12 e^{-\pi t^2}\,\text{erfi}(\sqrt{\pi}\,t)
	=\frac{1}{\sqrt\pi}\;2e^{-\pi t^2}\int_{0}^{\sqrt{\pi}t}e^{s^2}\,ds.
	\]
	Therefore, 
	\begin{equation}
		\hat g(t)
		=\frac{2}{\sqrt\pi}\,e^{-\pi t^2}
		\int_{0}^{\sqrt{\pi}t}e^{s^2}\,ds
		=\frac{2}{\sqrt\pi}\;\mathbb{D}_{+}\!\bigl(\sqrt{\pi}\,t\bigr),
	\end{equation}
	which is the desired HT of the GP $g(t)$.
	
\end{proof}
{\em Remarks:} (i) To verify orthogonality, consider 
\begin{equation}
	\int_{-\infty}^{\infty} g(t) \hat{g}(t)\,dt = \frac{2}{\sqrt{\pi}} \int_{-\infty}^{\infty} e^{-\pi t^2} \mathbb{D}_{+}(\sqrt{\pi} t)\,dt.
\end{equation}
Since $e^{-\pi t^2}$ is even and $\mathbb{D}_{+}(\sqrt{\pi} t)$ is odd, their product is odd. Therefore, 
\begin{equation}
	\int_{-\infty}^{\infty} g(t) \hat{g}(t) dt = 0.
\end{equation}
The zero inner product of the integral confirms that the GP and its HT are orthogonal.

(ii) The analytic signal is defined as
\begin{equation}
	g(t) + j \hat{g}(t) = \exp(-\pi t^2) + j \frac{2}{\sqrt{\pi}} \mathbb{D}_{+}(\sqrt{\pi} t). 
\end{equation}

Below, I present the result on the HT of $\GR_{I}(t ; T_0)$ and its proof, which is straightforward after applying the previous result on HT of $g(t)$.

\begin{result}
	\label{result:GRone_ht}
	Let $\widehat{\GR}_{I}(t; T_0)$ denote the HT of $\GR_{I}(t; T_0)$. The expression for $\widehat{\GR}_{I}(t; T_0)$ is given by
	\begin{equation}
		\widehat{\GR}_{I}(t; T_0)	=\sqrt{\frac{2}{\pi}}\Bigl[D_{+}(\sqrt\pi\,t)\;-\;D_{+}\bigl(\sqrt\pi\,(t-T_0)\bigr)\Bigr].
	\end{equation}
\end{result}
\begin{proof}
	I use the following proposition to prove the result.
	Let $g(t)$ be a real‐valued signal with Hilbert transform $\hat g(t)$. For any fixed delay $T_0$,
	\begin{equation}
		\mathcal{H}\{\,g(t) - g(t - T_0)\}(t)
		= \hat g(t)\;-\;\hat g(t - T_0),
	\end{equation}
	where $\mathcal{H}$ denotes the HT.
	
	Since \(\mathcal H\{g\}(t)=\tfrac{2}{\sqrt\pi}	\mathbb{D}_{+}(\sqrt\pi\,t)\), using the above proposition, I see that
	\begin{equation}
		\widehat{GR}_{I}(t; T_0)
		=\frac{2}{\sqrt{2\pi}}\Bigl[	\mathbb{D}_{+}(\sqrt\pi\,t)\;-\;	\mathbb{D}_{+}\bigl(\sqrt\pi\,(t-T_0)\bigr)\Bigr],
	\end{equation}
	where the Dawson function is
	\begin{equation}
		\mathbb{D}_{+}(s)=e^{-s^2}\int_0^s e^{v^2}\,dv.
	\end{equation}
\end{proof}

{\em Remarks:} (i) {\em Orthogonality:}

\begin{figure}[h!]
	\centerline{    \includegraphics[width=1\linewidth]{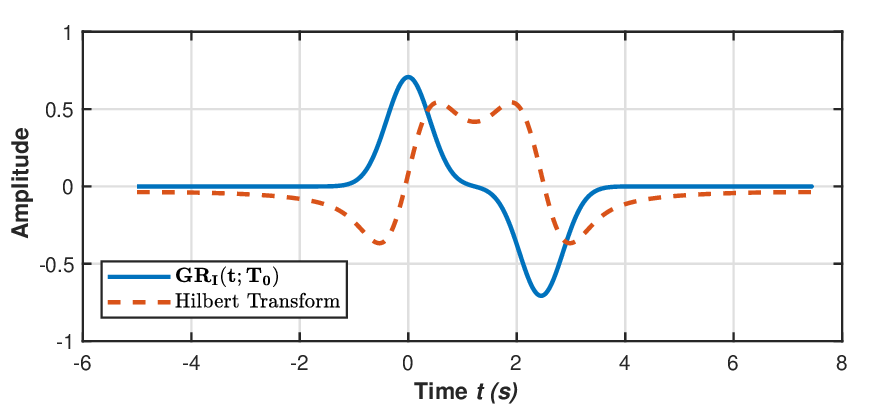}} 
	\caption{$\widehat{\GR}_{I}(t; T_0)$ and its Hilbert transform ($T_0 = 2.45$.).}
	\label{fig:GRFone_HilTrans_v1}
\end{figure}

Fig.~\ref{fig:GRFone_HilTrans_v1} illustrates the first-order GauRam function and its Hilbert transform. Their inner product is $\approx 0$ (precisely, $2.72 \times 10^{-17}$.).

(ii) {\em Analytic signal:} The analytic signal is given by $\GR_{I}(t; T_0) + j \widehat{\GR}(t ; T_0)$, which is useful to represent single sideband modulated signals. I present the details on Continuous-wave Gauss-Ramanujan modulation in the following section.

\section{APPLICATIONS OF GAURAM FUNCTIONS}

I consider some applications of GauRam functions in two important domains: communication systems and wavelets. First, I propose the continuous-wave (CW) Gauss-Ramanujan Modulation scheme and present its analysis. Later, I propose Gauss-Ramanujan shift keying and its analysis.

\subsection{Gauss-Ramanujan Modulation Scheme}

Constant--envelope, continuous--phase modulated signals are of high interest in modern wireless communications. The modulated signal design lies in efficiently using power amplifiers and achieving good spectral efficiency. 

Consider the transmission bandpass signal, which requires frequency translation. For it, consider the radio frequency carrier $\cos(2\pi f_c t)$ and its Hilbert transform $\sin(2\pi f_c t)$. Here, $f_c$ denotes the operating carrier frequency. Let $\GRM_{I}(t; f_c, T_0)$ denote the bandpass signal in the time domain. The In-phase and quadrature representation (or canonical representation) of the modulated signal is given by
\begin{equation}
	\label{eq:GRM_I_bp_td}
	\GRM_{I}(t; f_c, T_0) = \frac{1}{\sqrt{2}} e^{-\pi t^2} \cos(2\pi f_c t) 
	 - \frac{1}{\sqrt{2}}  e^{-\pi \brac{t-T_0}^2} \sin(2\pi f_c t).
\end{equation}

I express the bandpass signal in a complex envelope form by defining
\begin{align}
	I(t) &= \frac{1}{\sqrt{2}}\,e^{-\pi t^2}, \label{eq:I}\\[1mm]
	Q(t) &= -\frac{1}{\sqrt{2}}\,e^{-\pi (t-T_0)^2}. \label{eq:Q}
\end{align}
Therefore, the bandpass signal can be expressed as
\begin{equation}
	\GRM_{I}(t; f_c, T_0) = \Re\Bigl\{\Bigl[I(t) + j\,Q(t)\Bigr] e^{j2\pi f_c t}\Bigr\}.
	\label{eq:complex_representation}
\end{equation}

\begin{figure}[h!]
	\centerline{    \includegraphics[width=1\linewidth]{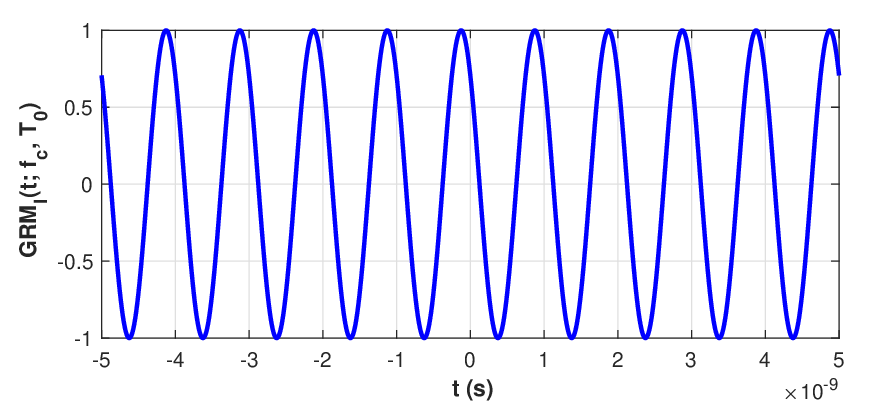}} 
	\caption{Modulated waveform  $\GRM_{I}(t; f_c, T_0)$ ($T_0 = 1.8$~ns, $f_c = 1$~GHz).}
	\label{fig:GRM_formOne_Illu_v1}
\end{figure}

A representative plot of the GRM waveform is shown in figure~\ref{fig:GRM_formOne_Illu_v1}.

{\em Remarks:} i) Using the standard CTFT pairs, I find that the frequency-domain representation of $\GRM_{I}(t; f_c, T_0)$ is given by
\begin{equation}
	\label{eq:GRM_I_bp_fd}
	\GRM_{I}(f; f_c, T_0) = \frac{1}{2\sqrt{2}} \brac{e^{-\pi (f - f_c)^2} + e^{-\pi (f + f_c)^2}}
	 - \frac{1}{2\sqrt{2}} e^{-j(2\pi f T_0 - \frac{\pi}{2})} \brac{e^{-\pi (f - f_c)^2} - e^{-\pi (f + f_c)^2}}.
\end{equation}

ii) The bandpass signal in~\eqref{eq:GRM_I_bp_td} is one possible  modulated waveform representation. However, practical modulation schemes are indeed possible. For instance, consider the truncated GP as in Gaussian-filtered minimum shift keying (GMSK)~\cite{haykin2014DigComSys}.

\subsubsection{Modulation Index Analysis}

For phase modulated or frequency modulated signals, the modulation index is often defined as the maximum phase deviation from the operating radio frequency carrier. The instantaneous phase of the complex envelope is given by
\begin{equation}
	\phi(t) = \arg\{I(t)+j\,Q(t)\}.
	\label{eq:phi_def}
\end{equation}

Substituting~\eqref{eq:I} and~\eqref{eq:Q}, I get
\begin{align}
	\phi(t) &= -\tan^{-1}\!\left(\frac{e^{-\pi (t-T_0)^2}}{e^{-\pi t^2}}\right), \\
	&= -\tan^{-1}\!\left(e^{\pi\bigl[t^2 - (t-T_0)^2\bigr]}\right).
	\label{eq:phi_calc}
\end{align}
%
Simplifying further, I find that the instantaneous phase becomes
\begin{equation}
	\phi(t) = -\arctan\!\Bigl(e^{\pi(2T_0 t-T_0^2)}\Bigr).
	\label{eq:phi_final}
\end{equation}

A natural definition for the modulation index \(m\) is the maximum absolute phase deviation. The more the pulses overlap, the more significant the phase deviation will be. At $t = \frac{T_0}{2}$,
\begin{equation}
	\phi\Bigl(\frac{T_0}{2}\Bigr) = -\arctan(1) = -\frac{\pi}{4}.
	\label{eq:phi_peak}
\end{equation}
I define the effective modulation index as
\begin{equation}
	m_{\rm GR} \triangleq \max_t |\phi(t)| \approx \frac{\pi}{4}\quad \text{(i.e. }45^\circ\text{)}.
	\label{eq:modulation_index}
\end{equation}

\subsubsection{Energy and Average Power Analysis}

Since the carrier modulation does not change the power, I evaluate the power of the complex envelope. Consider
\begin{equation}
	I^2(t) + Q^2(t) = \frac{1}{2}\left(e^{-2\pi t^2} + e^{-2\pi (t-T_0)^2}\right).
	\label{eq:envelope}
\end{equation}

The energy in one pulse is
\begin{align}
	E &= \int_{-\infty}^{\infty}\bigl[I^2(t) + Q^2(t)\bigr]\,dt \nonumber\\[1mm]
	&= \frac{1}{2}\left[\int_{-\infty}^{\infty}e^{-2\pi t^2}\,dt + \int_{-\infty}^{\infty}e^{-2\pi (t-T_0)^2}\,dt\right].
	\label{eq:energy_integrals}
\end{align}

Using the standard Gaussian integral
\begin{equation}
\int_{-\infty}^{\infty}e^{-\rho t^2}\,dt=\sqrt{\frac{\pi}{\rho}},
\end{equation}
I find that the energy is given by
\begin{equation}
	E = \frac{1}{2}\left(\frac{1}{\sqrt{2}} + \frac{1}{\sqrt{2}}\right) = \frac{1}{\sqrt{2}}.
	\label{eq:energy_final}
\end{equation}

If pulses are transmitted periodically with a repetition period $T_s$, the average power is given by
\begin{equation}
	P_{\rm avg} = \frac{E}{T_s} = \frac{1}{\sqrt{2}\,T_s}.
	\label{eq:avg_power}
\end{equation}

\subsubsection{Bandwidth Analysis}

Because the GRM waveform is constructed from GPs, its spectral characteristics follow the GP signal. Note that the Fourier transform of the GP $\exp\brac{-\pi t^{2}}$ is itself (in frequency domain). 

The $3$~dB point is defined by
\begin{equation}
	e^{-\pi f_{3dB}^2} = \frac{1}{\sqrt{2}},
	\label{eq:3dB_condition}
\end{equation}
which implies
\begin{equation}
	\pi f_{3dB}^2 = \frac{1}{2}\ln 2,\quad \text{or} \quad f_{3dB} = \sqrt{\frac{\ln2}{2\pi}}.
	\label{eq:f3dB}
\end{equation}
Because the Gaussian is symmetric, the full-width at half-maximum (FWHM) is
\begin{equation}
	\text{FWHM} = 2 f_{3dB} = \sqrt{\frac{2\ln2}{\pi}}.
	\label{eq:FWHM}
\end{equation}

For a time–scaled GP, for example, replacing \(t\) with \(t/T\) (so that the pulse becomes \(e^{-\pi (t/T)^2}\)), the spectral width scales inversely with \(T\). Hence, the effective bandwidth is
\begin{equation}
	BW \approx \frac{1}{T}\sqrt{\frac{2\ln2}{\pi}}.
	\label{eq:BW}
\end{equation}

Since the modulated GR signal in~\eqref{eq:GRM_I_bp_td} consists of two components (one modulated by $\cos(2\pi f_c t)$ and the other by $\sin(2\pi f_c t)$), the overall passband spectrum contains two Gaussian lobes centered at $\pm f_c$, each with approximately the same bandwidth as in~\eqref{eq:BW}.

%
%
%
%

\subsection{Gauss-Ramanujan Shift Keying (GRSK)}

Constant-envelope modulation schemes with continuous phase are highly desirable in modern wireless communication systems due to their power efficiency and spectral compactness. GMSK is a widely used example. Analogously, I propose Gauss--Ramanujan Shift Keying (GRSK), where the traditional Gaussian frequency pulse is replaced by a Gauss--Ramanujan frequency pulse. The extra design parameter inherent in the GR pulse provides additional flexibility for spectral shaping and interference mitigation.

\subsubsection{Waveform Design and Analysis}

For a binary data stream $\{d_{k}\}$ with $d_{k}\in\{+1,-1\}$ and modulation index $h$, a general single $h$-continuous-phase modulation (CPM) signal defined over $nT \leq t \leq (n+1)T$ is written as~\cite{Proakis_book}
\begin{equation}
	s(t) = \sqrt{\frac{2E}{T}}\,\cos\Bigl(2\pi f_c t + 2\pi h \sum_{k = -\infty}^{n} d_{k}\,q(t-kT)\Bigr),
	\label{eq:CPM}
\end{equation}
where $f_c$ is the operating carrier frequency,
$T$ is the symbol duration,
$q(t)$ is the waveform obtained by integrating a normalized frequency pulse \(g(t)\). The modulation index $h=0.5$ for minimum shift keying is a special case of continuous-phase frequency shift keying (CPFSK).

In GMSK, the signal pulse whose shape impacts the spectrum of the transmitted waveform is a GP defined over a finite interval~\cite{goldsmith_book}. Analogously, I define finite real-valued GR pulse as
\begin{equation}
	g_{\text{GR}}(t;T_0)= 
	\begin{cases}
		\displaystyle \frac{\kappa}{\sqrt{2}}\Bigl(e^{-\pi t^2}-e^{-\pi (t-T_0)^2}\Bigr), & 0\le t\le T, \\[2mm]
		0, & \text{otherwise}.
	\end{cases}
	\label{eq:final_GR_pulse_corrected}
\end{equation}
Here $T > T_0$ and $\kappa$ denote the normalization constant. The waveform $q(t)$ is defined as
\begin{equation}
	q_{\GR}(t)=\int_{0}^{t} g_{\text{GR}}(\tau;T_0)\,d\tau, \quad 0\le t\le T.
	\label{eq:raw_qt_integration}
\end{equation}

Using the standard integral~\cite{abramowitz_stegun}
\begin{equation}
	\int_{0}^{t} e^{-\pi \tau^2}\,d\tau=\frac{1}{2}\operatorname{erf}(\sqrt{\pi}\,t),
\end{equation}
I obtain, for \(0\le t\le T\),
\begin{equation}
	q_{\GR}(t)=\frac{\kappa}{2\sqrt{2}}\Bigl[\operatorname{erf}(\sqrt{\pi}\,t)-\operatorname{erf}\bigl(\sqrt{\pi}(t-T_0)\bigr)\Bigr].
	\label{eq:raw_qt}
\end{equation}
Here, $\operatorname{erf}$ denotes the error function.

%
The normalized phase waveform is given by
\begin{equation}
	\tilde{q}_{\GR}(t)=
	\begin{cases}
		0, & t<0,\\[1mm]
		\displaystyle \frac{1}{2}, & 0\le t\le T,\\[2mm]
		\frac{1}{2}, & t>T.
	\end{cases}
	\label{eq:final_phase_pulse_corrected}
\end{equation}

Thus, the proposed modulated GRSK waveform is given by
\begin{equation}
	s_{\text{GRSK}}(t)= \sqrt{\frac{2E}{T}}\,\cos\!\left(2\pi f_c t + 2\pi h \sum_{k = -\infty}^{n} d_{k}\,q_{\GR}(t-kT)\right).
	\label{eq:final_GRSK}
\end{equation}
This formulation guarantees a constant envelope and continuous phase transitions.

{\em Remarks on Continuous-Phase and Constant Envelope:} The instantaneous phase
\[
\phi(t)=2\pi f_c t+ 2\pi h \sum_{\forall k} d_{k}\,q_{\GR}(t-kT)
\]
is continuous because $q_{\GR}(t)$ is a smooth function defined in~\eqref{eq:final_phase_pulse_corrected}. Moreover, the cosine formulation in~\eqref{eq:final_GRSK} ensures a constant envelope essential for the efficient operation of power amplifiers.

{\em Remarks on Role of the Parameter $T_0$}
The delay parameter $T_0$ in~\eqref{eq:final_GR_pulse_corrected} provides additional control over the pulse shape. By adjusting $T_0$, one can tune the relative contributions of the two Gaussian components, thereby shaping the phase trajectory $q(t)$ to reduce inter-symbol interference (ISI) and better meet spectral mask requirements.

%

%
%

\subsubsection{Comparisons with GMSK}

In GMSK, a Gaussian filter is applied as a baseband frequency pulse, and its integral defines the phase pulse. GMSK and GRSK yield constant-envelope signals with continuous phase and produce passband spectra with two Gaussian-shaped lobes centered at \(\pm f_c\). The additional parameter \(T_0\) in GRSK allows further flexibility to optimize the phase response and sidelobe levels. With the proper choice of \(T_0\), GRSK can achieve improved spectral containment and reduced adjacent-channel interference compared to conventional GMSK.

{\em Remarks on power and bandwidth efficiency:}

i) Both GRSK and GMSK are constant-envelope schemes, ensuring efficient use of transmitted power and compatibility with nonlinear power amplifiers. The energy per symbol remains \(E\), and the amplitude of the signal in~\eqref{eq:final_GRSK} is $\sqrt{\frac{2E}{T}}$.

ii) The baseband frequency pulse mainly determines the occupied bandwidth. For GRSK, the effective $3$~dB bandwidth is approximately $\Delta f\approx \frac{1}{T}\sqrt{\frac{2\ln2}{\pi}}$. Although both GRSK and GMSK can be designed to have similar $3$~dB bandwidths, the extra flexibility provided by $T_0$ in GRSK can be exploited to shape the spectral sidelobes further and enhance spectral mask compliance. Below, I define the generalized form of the GRSK pulse waveform and evaluate its normalized PSD.

{\em A generalized form for GRSK pulses:} 

\emph{Euler’s totient function}, denoted by $\varphi(n)$ is the number of positive integers not exceeding $n$ that are relatively prime to $n$, that is, that satisfy $\gcd(k,n)=1\,$.

The $k$-th order GRSK pulse is defined as:
\begin{equation}
	\text{GR}_{k}(t ; T_0) = \sum_{n=0}^{\varphi(k) -1} \frac{r_k(n)}{\lVert r_k \rVert_2} \frac{\sqrt{\pi}}{\eta} e^{-\pi \frac{(t - nT_0)^2}{\eta^2}},
\end{equation}
where $\|.\|^2$ denote the $\mathcal{L}^{2}$ norm, $r_{k}(n)$ is the $n$-th Ramanujan sum coefficient, $\phi(k)$ is Euler's totient function, $T_0$ is the GP separation, and $\eta$ is the time-bandwidth product parameter. 

\subsubsection{GMSK vs. GPSK pulses: PSD Comparison}

Consider the first-order GMSK pulse given by
\begin{equation}
	\label{eq:GMSK}
	g_{\mathrm{GMSK}}(t)
	= \frac{\sqrt{\pi}}{\rho}\exp\!\Bigl(-\pi\,\frac{t^2}{\rho^2}\Bigr),
\end{equation}
where $\rho$ is the $3$\,dB bandwidth–time product.

Similarly, the GRSK pulse is given by
\begin{equation}
	g_{\mathrm{GRSK}}(t)
	=\frac{1}{\sqrt{2}}\frac{\sqrt{\pi}}{\eta}
	\exp\!\Bigl(-\pi\,\frac{t^2}{\eta^2}\Bigr)
	-\frac{1}{\sqrt{2}}\frac{\sqrt{\pi}}{\eta}
	\exp\!\Bigl(-\pi\,\frac{(t-T_0)^2}{\eta^2}\Bigr),
\end{equation}
%
%
with $\eta$ the $3$~dB bandwidth–time product. 

{\em Normalized PSD:} Let ${\mathrm{PSD}}_{n}(f)$ denote the normalized PSD. 

\begin{equation}
	\label{eq:norm_PSD}
	{\mathrm{PSD}}_{n}(f)
	= 10\log_{10}\!\Bigl(\frac{|G(f)|^2}{\max_f |G(f)|^2}\Bigr)\,\mathrm{dB},
\end{equation}
where $G(f)$ denotes the continuous-time Fourier transform of $	g_{\mathrm{GRSK}}(t)$ or $	g_{\mathrm{GMSK}}(t)$.


%
\begin{figure}[h!]
	\centerline{    \includegraphics[width=1\linewidth]{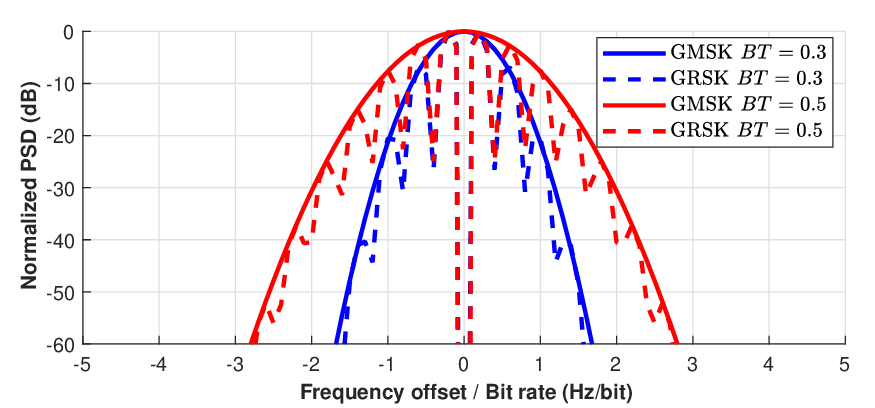}} 
	\caption{GMSK vs. GRSK for different bandwidth-time product ($BT$) values ($T_0 = 2.45$, $\rho = \frac{\sqrt{\ln 2}}{\pi BT}$.).}
	\label{fig:GMSK_vs_GRSK_v1}
\end{figure}

Fig.~\ref{fig:GMSK_vs_GRSK_v1} shows a representative plot that compares GRSK and GMSK pulse shapes. The orthogonal-like structure of GRSK pulses supports delay diversity and introduces a frequency-domain trade-off: subtracting two time-shifted Gaussians broadens the main lobe and raises the sidelobe floor. It creates deep, deterministic nulls at harmonics of $\frac{1}{T_0}$. This structured nulling capability enables precise notching of specific subbands without complex filter banks. Furthermore, the predictable sidelobe pattern can be further tuned via the choice of delay parameter~$T_0$ and Ramanujan sum weights, making GRSK a versatile spectral-shaping tool.

In Orthogonal Time Frequency Space (OTFS) modulation and other 5G-beyond systems, these spectral nulls translate into a sparse comb structure in the delay–Doppler domain, improving channel estimation and symbol recovery in highly dispersive channels.

\subsection{Gauss-Ramanujan Wavelets}

GP has superior time-frequency localization~\cite{Matz2007TWC}. To construct a mother Gauss-Ramanujan wavelet $\psi(t)$, I start with
\begin{equation}
	\nonumber
	\GR_{I}(t; T_0) = \frac{1}{\sqrt{2}}\,e^{-\pi t^2} - \frac{1}{\sqrt{2}}\,e^{-\pi (t-T_0)^2}.
\end{equation}

A function $\psi(t)$ qualifies as a mother wavelet if it satisfies the following conditions: i) {\em Zero Mean (admissibility)~\cite{Burrus1997book}:} Mathematically,
\begin{equation}
	\label{eq:AdmCon}
	\int_{-\infty}^{\infty} \psi(t) \, dt = 0.
\end{equation}

Since the integral of the GP is invariant under translation, it is easy to show that the zero-mean condition is met.

ii) {\em Unit Energy:}
Mathematically, 
\begin{equation}
	\label{eq:UnitEng_MW}
	\int_{-\infty}^{\infty} \abs{\psi(t)}^2\, dt = 1.
\end{equation}

Since  $\int_{-\infty}^{\infty} |\GR_{I}(t; T_0)|^2\, dt \neq 1$, appropriate normalization is needed to construct the mother wavelet. Below, I present the details on the admissibility condition and normalization.

\subsubsection{Admissibility Condition}

As mentioned before, the admissibility condition requires that the mother wavelet has zero mean. I find that
\begin{equation}
	\int_{-\infty}^{\infty} 	\GR_{I}(t; T_0)\, dt = 0.
\end{equation}
Since the integral of a GP is independent of its center, I have
\begin{equation}
	\label{eq:Adm_cond}
	\int_{-\infty}^{\infty} e^{-\pi t^2}\, dt = \int_{-\infty}^{\infty} e^{-\pi (t-T_0)^2}\, dt = I.
\end{equation}
Further simplification yields
\begin{equation}
	\label{eq:Adm_Cond_simp}
\int_{-\infty}^{\infty} \GR_{I}(t; T_0)\, dt = \frac{1}{\sqrt{2}}I - \frac{1}{\sqrt{2}}I = 0.
\end{equation}

Therefore, the proposed function satisfies the zero-mean requirement. Below, I verify the normalization requirement.

\subsubsection{Normalization to Unit Energy}

In addition to zero mean, the mother wavelet must be normalized such that its energy is unity:
\begin{equation}
	\|\GR_{I}(t; T_0)\|^2 = \int_{-\infty}^{\infty} \left|\GR_{I}(t; T_0)\right|^2 dt = 1.
\end{equation}

A direct computation shows that the energy of the function $\GR_{I}(t; T_0)$ is given by
\begin{equation}
	\label{eq:norm_GP}
\|\GR_{I}(t; T_0)\|^2 = \frac{1}{2} \times \\ \left[\|e^{-\pi t^2}\|^2 + \|e^{-\pi (t-T_0)^2}\|^2 - 2\langle e^{-\pi t^2},\, e^{-\pi (t-T_0)^2}\rangle\right],
\end{equation}
where $\|.\|^2$ denote the $\mathcal{L}^{2}$ norm and $\langle . \rangle$ denotes the inner product.

Because the cross product term is nonzero, the initial scaling does not guarantee unit energy. Therefore, I introduce a normalization constant $\mathcal{B}$ and redefine the wavelet as
\begin{equation}
	\label{eq:Mother_wl}
	\psi_{\text{GR}}(t) = \mathcal{B}\Bigl[e^{-\pi t^2} - e^{-\pi (t-T_0)^2}\Bigr],
\end{equation}
where $\mathcal{B}$ is determined by the normalization condition. I find that
\begin{equation}
	\label{eq:Norm_factor}
\mathcal{B} = \frac{1}{\sqrt{\sqrt{2}\,\Bigl(1 - e^{-\frac{\pi T_0^2}{2}}\Bigr)}}.
\end{equation}
This normalization ensures that $\psi_{\text{GR}}(t)$ meets both the zero-mean and unit-energy conditions required for a valid mother wavelet.

{\em Remarks:}

The novel Gauss-Ramanujan wavelet $\psi_{\text{GR}}(t)$ exhibits desirable properties such as smoothness, excellent time-frequency localization, and inherent symmetry. These attributes make it a suitable tool for a variety of signal processing applications, including:

\begin{itemize}
	\item {\em Time-Frequency Analysis:} The wavelet transform based on $\psi_{\text{GR}}(t)$ yields a detailed representation of signal components in both time and frequency domains, facilitating the analysis of transient phenomena.
	
	\item {\em Denoising:} In wavelet-based denoising techniques, the localization properties of $\psi_{\text{GR}}(t)$ allow for efficient discrimination between signal features and noise.
	
	\item {\em Feature Extraction:} The wavelet transform can highlight essential features in signals, proving useful in pattern recognition and machine learning applications.
\end{itemize}

\subsubsection{Gauss-Ramanujan wavelet versus Hermite wavelet: Benchmarking}

I consider two unit‐energy wavelets, namely, $\psi_{\text{H}}(t)$ and $\psi_{\GR}(t)$. The former represents the first-order Hermite wavelet, and the latter represents the first-order Gauss-Ramanujan wavelet.
%
%
\begin{align}
	\psi_{\text{H}}(t) &= \sqrt{2}\,\pi^{-1/4}\;t\;e^{-t^2/2}, 
	\label{eq:hermite}\\
	\psi_{\GR}(t) &= \mathcal B\Bigl[e^{-\pi t^2} - e^{-\pi(t - T_0)^2}\Bigr].
	\quad
	\label{eq:norm}
\end{align}

{\em The autocorrelation functions:} Using \(\psi_H\) and the definition of autocorrelation function, I get
\begin{equation}
	R_{\text{H}}(\tau) 
	= \Bigl(1 - \tfrac{\tau^2}{2}\Bigr)\,e^{-\tau^2/4}.
	\label{eq:R_H}
\end{equation}

Similarly, using $\psi_{\GR}(t)$ and the definition of autocorrelation function, I get
\begin{equation}
	R_{\GR}(\tau)
	= \frac{2\,e^{-\frac{\pi}{2}\tau^2} 
		- e^{-\frac{\pi}{2}(\tau - T_0)^2}
		- e^{-\frac{\pi}{2}(\tau + T_0)^2}}
	{2\bigl(1 - e^{-\pi T_0^2/2}\bigr)},
	\label{eq:R_GR}
\end{equation}
\begin{figure}[h!]
	\centerline{    \includegraphics[width=1\linewidth]{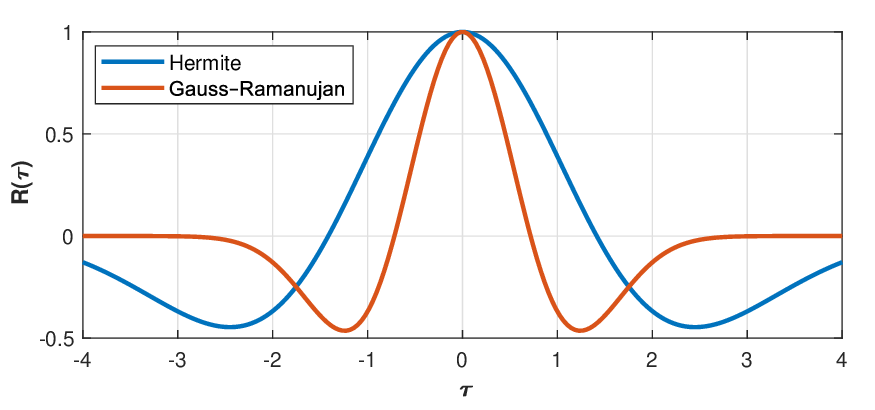}} 
	\caption{Comparison of Hermite and Gauss-Ramanujan wavelets of the first order ($T_0 = 1$.).}
	\label{fig:Hermite_vs_GR_v1}
\end{figure}

Fig.~\ref{fig:Hermite_vs_GR_v1} shows autocorrelation functions of Hermite and Gauss–Ramanujan wavelets (\(T_0=1\)). The rapid decay of the Gauss-Ramanujan wavelet indicates tighter time‐containment.

{Gauss-Ramanujan wavelet versus Hermite wavelet: comparison of time-frequency containment:} 

I quantify joint time-frequency localization by the variances as follows.
\begin{equation}
	\Delta t^2 \;=\;\int_{-\infty}^{\infty} t^2\,\bigl|\psi(t)\bigr|^2\,dt,
	\quad
	\Delta \omega^2 \;=\;\int_{-\infty}^{\infty} \omega^2\,\bigl|\Psi(\omega)\bigr|^2\,\frac{d\omega}{2\pi},
\end{equation}

I form the uncertainty product \(\Delta t\,\Delta\omega\).  Now, I evaluate these integrals numerically. Let \(T_0 = 1\).
\begin{table}[!htb]
	\caption{Time–Frequency Containment of \(\psi_{\text{H}}\) vs.\ \(\psi_{\GR}\) (\(T_0=1\))}  
	\label{tab:tf_compare}
	\centering
	\begin{tabular}{lccc}
		\hline
		Wavelet & \(\Delta t\) & \(\Delta \omega\) & \(\Delta t\,\Delta \omega\) \\
		\hline
		Hermite\,1 & \(\sqrt{1.5}\approx1.225\) & \(\sqrt{0.5}\approx0.707\) & \(0.866\) \\
		Gauss–Ramanujan & \(0.720\) & \(1.055\) & \(0.760\) \\
		\hline
	\end{tabular}
\end{table}

From Table~\ref{tab:tf_compare}, I state the following remarks:

\begin{itemize}
	\item {\em Superior localization:}  The GR wavelet has a smaller uncertainty product (\(0.760\)) than the Hermite wavelet (\(0.866\)), indicating tighter joint time–frequency containment.
	
	\item {\em Adaptable trade‐off:}  The shift \(T_0\) in \(\psi_{\GR}\) allows continuous adjustment of \(\Delta t\) and \(\Delta\omega\).  In contrast, Hermite‐order wavelets have fixed spreads determined by their polynomial order.
	
	\item {\em Use cases:}  In use cases such as denoising and pulse detection, the improved time-frequency containment of \(\psi_{\GR}\) yields efficient impulse responses, higher peak‐to‐sidelobe ratios, and lower reconstruction error. Detailed analysis and validation of a specific application could be potential future work.
\end{itemize}

\subsection{Additional remarks on features and applications}

The salient features and applications of the proposed Gauss-Ramanujan wavelet and GRSK scheme are as follows:

\begin{itemize}
	\item \textbf{Efficient time-frequency containment:} The GauRam functions are designed to provide excellent localization in both time and frequency domains. This property directly supports energy-efficient transmission by minimizing spectral leakage and reducing out-of-band emissions, critical for power-constrained devices in IoT, sensor networks, and green wireless systems.
	
	\item \textbf{Adaptability via tunable delay parameter:} The proposed GauRam functions involve a delay parameter $T_0$ and Ramanujan weight structures that allow dynamic Adaptability to dynamic propagation environments (for example, multipath delay, Doppler spread). This tunability is particularly beneficial for hybrid systems (for example, THz/mmWave, RF/optical), where the waveform must adapt to heterogeneous propagation environments.
	
	\item \textbf{Near-orthogonality and interference resilience:} The near-orthogonal properties and low mean overlap reduce inter-symbol interference and enhance detection performance. These features are vital for densely packed communication scenarios and hybrid multi-input, multi-output (MIMO) systems where mutual interference is a limiting factor.
	
	\item \textbf{Low-complexity implementation:} The analytical simplicity of the first-order GauRam functions enables low-complexity implementations on hardware, reducing power consumption during real-time signal processing. This feature is aligned with the goals of green and energy-harvesting communication systems.
	
	\item \textbf{Hybrid use cases:} The continuous-time Gauss-Ramanujan Modulation (GRM) and the digital GRSK scheme can be integrated into systems that combine classical modulations (e.g., GMSK, OFDM) with emerging paradigms like OTFS, chirp spread spectrum (CSS), or delay-Doppler signaling. The inherent flexibility and robustness of the proposed schemes allow seamless transition across frequency bands and signaling formats.
	
	\item \textbf{Wavelet applications in signal processing:} In addition to communication applications, the Gauss-Ramanujan wavelet—owing to its improved Adaptability compared to Hermite wavelets—can be employed in low-power signal representation, denoising, and compressive sensing frameworks, contributing further to energy-efficient signal analysis and feature extraction in embedded systems.
\end{itemize}

\section{CONCLUSIONS AND FUTURE RESEARCH}

Motivated by Ramanujan sums and the significant properties of GP, I developed and presented novel constructions called GauRam functions. These are based on sequences involving Ramanujan sums, GP, and DGP. I developed insightful analysis for deterministic and stochastic overlap between GP and DGP. Specifically, I presented exact and closed form approximation expressions for mean overlap and evaluated them. Later, I presented a procedure for constructing GauRam functions and derived mathematical (spectral) properties of selected GauRam functions. I extended the analysis and analyzed HT of the first-order GauRam functions to obtain more useful insights into analytic signal representations. Finally, I suggested their use in communication systems, specifically the Gauss-Ramanujan modulation scheme and Gauss-Ramanujan wavelet construction.
Furthermore, I presented insightful applications of these functions in modern communication systems and signal processing. Specifically, I present the continuous-wave GRM scheme, GRSK scheme, and Gauss-Ramanujan wavelets, and their analysis with comparisons and benchmarking. These modulation schemes and wavelets exhibit useful properties (for example, PSD and time-frequency containment), enabling their use in next-generation terrestrial and satellite communication systems and signal processing.

Interesting future avenues lie in the rigorous performance analysis of continuous-wave GRM and GRSK schemes in various communications and Gauss-Ramanujan wavelets. Specifically, the performance of modulation schemes in terms of the figure of merit (for CW modulation scheme) and symbol error probability (for digital modulation scheme) can be investigated.

\section*{ACKNOWLEDGMENT}

The author would like to thank BITS Pilani, Pilani Campus, for providing the laboratory facilities and software necessary for conducting this research.





%
%

\vfill\pagebreak


\begin{thebibliography}{00}
	
\bibitem{haykin2014DigComSys}
S.~Haykin, {\em Digital communication systems}, \newblock John Wiley \& Sons, 2014.

\bibitem{oppenheim1999discrete}
A.~V. Oppenheim and R.~W. Schafer, {\em Discrete-Time Signal Processing}, 2nd~ed.
\newblock Prentice Hall, 1999.

\bibitem{Atlas1999SPmag}
L. Atlas and P. Duhamel, "Recent developments in the core of digital signal processing," {\em IEEE Signal Proc. Mag.}, vol. 16, pp. 16--31, Jan. 1999.

\bibitem{hadani2017orthogonal}
R.~Hadani \textit{et al.}, ``Orthogonal time frequency space modulation,'' \textit{Proc. IEEE WCNC}, 2017, pp. 1--6.

\bibitem{raviteja2018interference}
P.~Raviteja, K.~T. Phan, Y.~Hong, and E.~Viterbo, ``Interference cancellation and iterative detection for orthogonal time frequency space modulation,'' \textit{IEEE Trans. Wireless Commun.}, vol. 17, no. 10, pp. 6501--6515, Oct. 2018.

\bibitem{surabhi2019optimal}
G. D. Surabhi, R. M. Augustine and A. Chockalingam, "Peak-to-average power ratio of OTFS modulation," {\em IEEE Communications Letters}, vol. 23, no. 6, pp. 999--1002, June 2019. 

\bibitem{Aldababsa2024otfs}
M. Aldababsa, S. Özyurt, G. K. Kurt and O. Kucur, "A Survey on orthogonal time frequency space modulation," {\em IEEE Open Journal of the Communications Society}, vol. 5, pp. 4483--4518, 2024.

\bibitem{mallat1999wavelet}
S.~Mallat, \textit{a Wavelet Tour of Signal Processing}. Academic Press, 1999.

\bibitem{duarte2011structured}
M.~F. Duarte and Y.~C. Eldar, ``Structured compressed sensing: from theory to applications,'' \textit{IEEE Trans. Signal Process.}, vol. 59, no. 9, pp. 4053--4085, Sep. 2011.

\bibitem{PPV2015Eusipco}
P. P. Vaidyanathan and S. Tenneti, "Properties of Ramanujan filter banks," in {\em Proc. {EUSIPCO}}, Dec. 2015.

\bibitem{Lopez2011ieeeTcom}
M. López-Benítez and F. Casadevall, "Versatile, accurate, and analytically tractable approximation for the Gaussian Q-function," {\em IEEE Trans.\ Commun.}, vol. 59, pp. 917--922, Apr. 2011.

\bibitem{bracewell}
R. N. Bracewell, \emph{The Fourier Transform and Its Applications}. 3rd~ed.\newblock McGraw-Hill, 1999.

\bibitem{boashash}
B. Boashash, \emph{Time-frequency signal analysis and processing$—$a comprehensive reference}. \newblock Elsevier Science, 2003.

\bibitem{King2009}
F.~W.~King, *Hilbert Transforms*, Encyclopedia of Mathematics and its Applications. Cambridge, U.K.: Cambridge University Press, 2009.


\bibitem{abramowitz_stegun}
M.~Abramowitz and I.~Stegun, {\em Handbook of mathematical functions with formulas, graphs, and mathematical tables}.
\newblock Dover, 9~ed., 1972.

\bibitem{Proakis_book}
J. G. Proakis, {\em Digital Communications}. \newblock McGraw-Hill, 2008.

\bibitem{goldsmith_book}
A.~J. Goldsmith, {\em Wireless Communications}.
\newblock Cambridge University Press, 2005.

\bibitem{Matz2007TWC}
G. Matz, D. Schafhuber, K. Grochenig, M. Hartmann and F. Hlawatsch, "Analysis, optimization, and implementation of low-interference wireless multicarrier systems," {\em IEEE Trans.\ Wireless
	Commun.}, vol. 6, pp. 1921--1931, May 2007.

\bibitem{Burrus1997book}
C. S. Burrus \emph{et al.}, {\em Introduction to Wavelets and Wavelet Transforms: A Primer}. Prentice Hall, 1998. 

\end{thebibliography}
\end{document}